# The Synthesis, Structure and Electronic Properties of a Lead-Free Hybrid Inorganic-Organic Double Perovskite (MA)2KBiCl6 (MA = methylammonium)

Fengxia Wei,[a] Zeyu Deng,[a] Shijing Sun,[a] Fei Xie,[a] Gregor Kieslich,[a] Donald M. Evans,[b] Michael A. Carpenter,[b] Paul D. Bristowe[a] and Anthony K. Cheetham[a]*

[a.]Department of Materials Science and Metallurgy, University of Cambridge, 27 Charles Babbage Road, CB3 0FS, United Kingdom

[b.]Department of Earth Sciences, University of Cambridge, S109 Downing Street, CB2 3EQ, United Kingdom

* Corresponding author: akc30@cam.ac.uk

ABSTRACT: IN A SEARCH FOR LEAD-FREE MATERIALS THAT COULD BE USED AS ALTERNATIVES TO THE HYBRID PEROVSKITES, (MA)PbX$_3$, in photovoltaic applications, we have discovered a hybrid double perovskite, (MA)$_2$KBiCl$_6$, which shows striking similarities to the lead analogues. Spectroscopic measurements and nanoindentation studies are combined with density functional calculations to reveal the properties of this interesting system.

The light-harvesting, semiconducting hybrid inorganic-organic perovskites (HIOPs) have recently attracted a great deal of attention in the photovoltaic community, with their solar cell efficiencies rising from ~4% to over 20% in just six years.[1, 2] The most extensively studied materials are the lead-containing systems, APbX$_3$, where A is alkyl ammonium cation (e.g. CH$_3$NH$_3^+$ (methylammonium, MA) or NH$_2$CHNH$_2^+$ (formamidinium, FA)) and X is Cl$^-$, Br$^-$ or I$^-$. However, the toxicity of lead to the environment could become a major drawback in their commercialization and the quest for lead-free alternatives is therefore attracting a lot of attention. Other group IV metals such as Ge and Sn are being explored, but the chemical instability of Sn$^{2+}$ and Ge$^{2+}$ presents challenges for their practical utilization.[3, 4] Alternatively, the replacement of Pb$^{2+}$ by isoelectronic ions also seems attractive because the strong light absorption and long carrier life-times exhibited by MAPbX$_3$ are believed to be related to the 6s$^2$6p$^0$ electronic configuration of Pb$^{2+}$.[5] While Tl$^+$ is also toxic, Bi$^{3+}$ is an interesting option because coordination complexes of bismuth are used in over-the-counter medicines such as Pepto-Bismol.[6] However, this strategy poses challenges because Bi$^{3+}$ has a different valence state from Pb$^{2+}$ and cannot therefore be simply substituted into phases such as (MA)PbX$_3$. In the present work, we show that the incorporation of Bi$^{3+}$ into a HIOP can be achieved by synthesizing a hybrid double perovskite of general formula (MA)$_2$M$^I$M$^{III}$X$_6$.

There has been significant recent progress in the incorporation of Bi$^{3+}$ into hybrid perovskite-related halides. For example, (MA)$_3$Bi$_2$I$_9$ can be readily obtained by using a synthetic route analogous to that used for MAPbI$_3$,[7] and an ammonium bismuth iodide phase, (NH$_4$)$_3$Bi$_2$I$_9$, was recently reported to show a bandgap of 2.04eV.[8] A number of alkali metal systems of composition $M^I_3Bi_2I_9$ (M = K, Rb, Cs) have also been described.[9, 10] None of these systems is a perovskite, however, and their structures have lower dimensionalities and therefore wider band gaps than their 3-D analogues. In order to maintain the 3D perovskite architecture, an alternative approach is to use a combination of a monovalent cation, such as an alkali metal or $Ag^+$, and a trivalent cation such as $Bi^{3+}$, in an ordered B-site arrangement to form a double perovskite structure of general formula $A^I_2B^IB^{III}X_6$. During the 1970s, a number of double perovskites (which were described by their alternative name, elpasolites) of composition $Cs_2NaM^{III}Cl_6$ ($M^{III}$ = lanthanide, actinide, bismuth, and so on) were reported in $Fm\bar{3}m$ symmetry, wherein the perovskite network contains alternating $NaCl_6$ and $M^{III}Cl_6$ octahedra on the B-sites.[11] At that time, particular interest focused on a ferroelectric phase transition that was observed on cooling $Cs_2NaBiCl_6$.[12-15] Very recently, and in the light of the intense interest in halide perovskites, McClure et. al. reported $Cs_2AgBiX_6$ (X = Cl, Br) with a bandgap of 2.77eV for X = Cl and 2.19eV for X = Br,[16] while a separate paper by Slavney et. al. described the X = Br phase with a bandgap of 1.95eV and a long photoluminescence lifetime of ca. 660ns.[17] The observation that hybrid perovskites are more easily processed into devices and show different optical properties compared with their inorganic analogues[18] encouraged us to investigate the possibility of making hybrid double perovskites. Here we report such a material, $(MA)_2KBiCl_6$, which crystallizes in rhombohedral $R\bar{3}m$ symmetry and contains alternating $KCl_6$ and $BiCl_6$ octahedra that are corner-sharing to form a 3-D network (Figure 1). The experimentally observed structure is compared with Density Functional Theory (DFT) calculations, and we also report the results of absorption spectroscopy, thermal analysis, resonant ultrasound spectroscopy (RUS) and nanoindentation measurements.

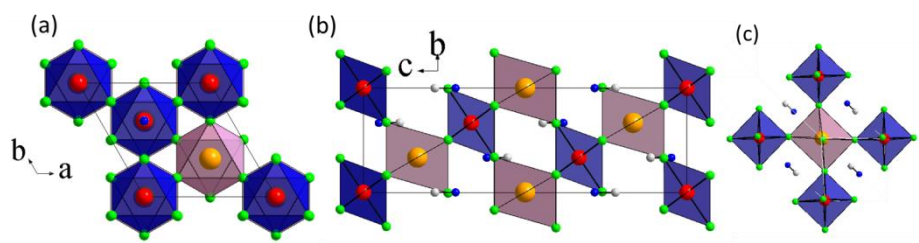

Figure 1. Crystal structure of $(MA)_2KBiCl_6$, obtained from single crystal X-ray diffraction, viewed along (a) the *c* axis, (b) the *a* axis, and (c) tilted to show the MA location. Red: Bi, brown: K, green: Cl, white: C, blue: N. $BiCl_6$ and $KCl_6$ octahedra are shown in blue and purple, respectively.

Synthesis: The starting material, $CH_3NH_3Cl$, was prepared by mixing stoichiometric amounts of methylamine solution (40wt% in $H_2O$) and HCl (37% in $H_2O$) at 0 °C, then heating at 60 °C to dryness, washing with acetone, and drying overnight in a vacuum oven. Crystals of $(MA)_2KBiCl_6$ were synthesized at 150 °C by the hydrothermal method in a stainless steel Parr autoclave using 2.4mmol $CH_3NH_3Cl$, 1.2mmol KCl and 1.2mmol $BiCl_3$ in 1ml HCl acid solution. Single phase powders can be obtained through solvent precipitation from hot HCl acid.

Thermogravimetric analysis (TGA) and differential scanning calorimetry (DSC) were conducted using an SDT (simultaneous DSC-TGA) Q600 instrument. 13.89mg of powder samples were heated from room temperature to 800 °C at 10 °C/min under a nitrogen flow of 100ml/min. A peak was observed at ca. 60 °C in the DSC curve, indicating a possible phase transition as there was no mass loss at the corresponding TGA curve (Figure S1). Samples started to decompose from ca. 230 °C, which is near the MACl boiling range (225 °C-230 °C) and BiCl$_3$ melting point (227 °C). Around 60% weight loss was observed at ca.430 °C, and the amount reached ~77% at ca. 563 °C.

Crystal structure determination was carried out using an Oxford Gemini E Ultra diffractometer, Mo Kα radiation (λ = 0.71073Å), equipped with an Eos CCD detector. Data collection and reduction were conducted using CrysAliPro (Agilent Technologies). An empirical absorption correction was applied with the Olex2 platform,[19] and the structure was solved using ShelXT[20] and refined by ShelXL.[21] Phase purity was confirmed by powder X-ray diffraction (PXRD, Bruker D8 diffractometer with primary Ni filter). Lattice parameters obtained from a Pawley fit performed using TOPAS4.1[22] are in good agreement with lattice parameter from single crystal X ray diffraction. No secondary phase was detected for products synthesized by both solution precipitation and hydrothermal methods (Figure S2).

(MA)$_2$KBiCl$_6$ crystallizes in $R\bar{3}m$ symmetry (CCDC 145389), with lattice parameters a = 7.8372(2)Å and c = 20.9938(7)Å. Alternating KCl$_6$ and BiCl$_6$ octahedra share corners to form an ReO$_3$ framework and the MA cations occupy the A-site cavities to complete the double perovskite structure. We estimate that (MA)$_2$KBiCl$_6$ has a Tolerance Factor of 0.93, which is well within the range (0.80-1.0) where the perovskite structure can be expected to form.[23] Due to the obvious size difference between the radii of K+ (1.38Å) and Bi3+ (1.03Å),[24] the 3D network is distorted from the cubic stacking that is seen in caesium-based Na/Bi and Ag/Bi double perovskites. The KCl$_6$ octahedra are slightly distorted with K-Cl bond lengths of 3.049(2)Å and two octahedral bond angles of 81.67(9) ° and 98.33(9) °, while the BiCl$_6$ octahedra are more regular in shape with Bi-Cl distances of 2.681(2)Å and bond angles of 91.76(10) ° and 88.24(10) ° (Tables S3 and S4). The bridging angle between the KCl$_6$ and BiCl$_6$ octahedra is slightly bent with a K-Cl-Bi angle of 173.04(12) °. The structure is less tilted than the low-temperature orthorhombic form of (MA)PbCl$_3$, where the Pb-Cl-Pb bond angles range from 154.88(11) to 169.46(4) °.[25]

The MA cations in the perovskite A-site are aligned along the c direction (Figure 1c) and alternate; symmetry-equivalent MA cations adopt opposite orientations (C-N ⋯ N-C) and have C-N bond lengths of 1.35(3)Å. The X-ray measurements are unable to locate the hydrogen atoms, so we cannot be sure whether the MA cations are ordered (which is allowed in space group $R\bar{3}m$), or disordered. However, hydrogen bonding is implied between the amine and the chloride anion since the N ⋯Cl distance (3.406(1)Å) is shorter than the C ⋯Cl one (3.848(2)Å). The corresponding distances for N…Cl in orthorhombic (MA)PbCl$_3$ are 3.273(2)Å, 3.346(2)Å, 3.366(10)Å and 3.421(2)Å,[25] suggesting that hydrogen bonding is important for both structures. Strong anisotropic atomic displacement parameters for the Cl anions indicate a possible rotational/tilting feature for the octahedra (Figure S3), which may be associated with MA cation disorder.

Optical measurements were carried out on a PerkinElmer Lambda 750 UV-Visible spectrometer in the reflectance mode with a 2nm slit width. The scan interval was 1 nm and the scan range was between 200 and 1500nm. The measurement was repeated by rotating the sample holder by 90 ° to ensure that the geometrical condition was fulfilled. The reflectance spectrum shows two edges and values of 3.04eV and 3.37eV were obtained for the apparent optical bandgaps from the corresponding Tauc plot (Figure 2). The DFT calculations (below) have allowed us to interpret these observations.

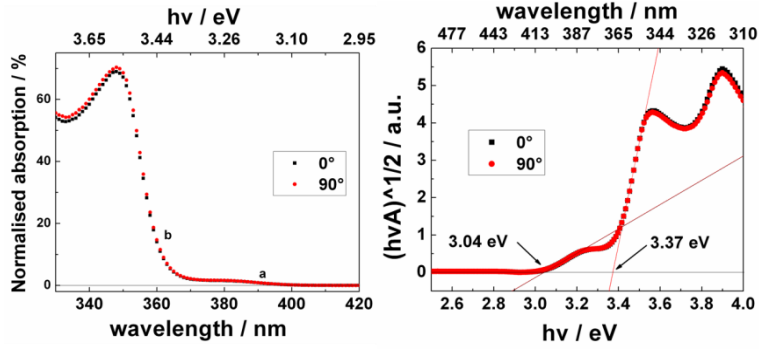

Figure 2. (a) Normalized absorption spectra converted from reflectance data and (b) the Tauc plot (assumed indirect bandgap).

Nanoindentation experiments were performed at room temperature using an MTS NanoIndenter XP (MTS Corp., Eden Prairie, MN) located in an isolation cabinet in order to minimize the thermal instability and acoustic interference. With a dynamic displacement-controlled continuous stiffness measurement (CSM mode), as reported in previous literature,[26, 27] indentations were carried out using a three-sided pyramidal Berkovich tip (radius ~100 nm) with the indenter aligned normal to the (001) plane of the single crystals. The Young's modulus (E) and indentation hardness (H) were obtained using the standard Oliver–Pharr Method.[28, 29] Calibration was done using a fused silica standard with an elastic modulus of 72 GPa and hardness of 9 GPa. Poisson's ratio of the samples was assumed to be 0.3 in this model. Sample preparations followed the similar procedures of Sun et. al..[30]

The Young's modulus was found to be 14.6 ± 0.3 GPa and the hardness 0.15 ± 0.03 GPa, averaging the results from an indentation depth of 200 nm to 900 nm (See Figure S5 and S6). In comparison with the experimental mechanical properties of the lead-based perovskites, the Young's modulus of $(MA)_2KBiCl_6$ is higher than that of $(MA)PbI_3$ (E ≈ 11 GPa) and lower than that of $(MA)PbCl_3$ (E ≈ 17-20 GPa).[30] Since the crystal structures of $(MA)PbCl_3$ and $(MA)_2KBiCl_6$ adopt the same topology and differ only in terms of the B-site cations, the greater compliance of $(MA)_2KBiCl_6$ can be attributed to the weaker K-Cl bonding in comparison with that of Pb-Cl. Similar mechanical behaviour has been observed in previous work. For example, replacing $Zn^{II}$ cations with alternating $Li^I$ and $B^{III}$ in zeolitic imidazolate frameworks (ZIFs) reduces the elastic modulus from 8 – 9 GPa for $Zn(Im)_2$ to ~3 GPa for $LiB(Im)_4$.[31]

DFT calculations were performed using the Vienna ab initio Simulation Package (VASP).[32,33] Projector-augmented wave (PAW)[34] pseudopentials were used with the following valence electrons for each ion treated explicitly: Bi ($5d^{10}6s^26p^3$), K ($3s^23p^64s^1$), Cl ($3s^23p^5$), C ($2s^22p^2$), N ($2s^22p^3$) and H ($1s^1$). The semi-local van der Waals functional (vdW-DF) was used,[35] with the exchange energy calculated using the optimized optB86b generalized gradient approximation (GGA) functional. The correlation energy was obtained from the local density approximation (LDA) and the nonlocal correlation energy calculated from double space integration. A 500 eV planewave kinetic energy cutoff was used for all the calculations. For geometry optimisation a 4×4×2 Monkhorst-Pack[36] k-point mesh was employed, while for the electronic density of states (DOS) calculations a finer 5×5×2 mesh was used. The ions were relaxed until the forces on them were less than 0.01 eV/Å. Relativistic spin-orbit coupling (SOC) was included in the DOS and band structure calculations.

DFT geometry optimization in $R\bar{3}m$ results in lattice parameters of a = 7.8165Å and c = 20.9904 Å, which agree well with the single crystal XRD measurements. The DFT methods were used to predict the locations of the hydrogen atoms which could not be determined using X-Rays (though they could be obtained using neutron diffraction[37]). Note that since the space group could be consistent with both ordered or disordered hydrogen positions, the calculations do not establish whether the system is ordered at room temperature. Tables S1, S3 and S4 compare the fractional atomic coordinates, bond lengths and bond angles obtained from the calculations with those from the X-ray measurements; again it is seen that the agreement is very good. The maximum deviation is in the C-N bond distance (1.492Å vs 1.35(3)Å), which corresponds to the hydrogen bonding interaction.

Figure 3(a,b) shows the calculated band structure of $(MA)_2KBiCl_6$, which predicts an indirect band gap of 3.08eV, in good agreement with our optical measurements (3.04eV). Note that a direct transition would also be possible at 3.20eV, which is consistent with the experimental feature at 3.37eV (see above). The corresponding total and partial electronic density of states (Figure S7(a)) reveal that MA does not contribute to states near the valence band maximum (VBM) or the conduction band minimum (CBM), which is similar to the case for $(MA)PbI_3$.[38] It is seen that Bi and Cl are the main contributors to states near the band gap. This can be further visualized in real space (Figure S7(b)) by calculating the band decomposed partial charge density (PCD). It is seen that the valence band edge is composed of Bi-6s and Cl-3p antibonding states whereas the conduction band edge is composed of Bi-6p, Cl-3p antibonding states together with a small contribution from Cl-3s. However, the energy states of K are low lying; e.g. the K-3p states (not shown in the figure) occur at ~13 eV below the VBM. This differs from $Cs_2AgBiBr_6$[16] where Ag contributes energy states near the valence band edge.

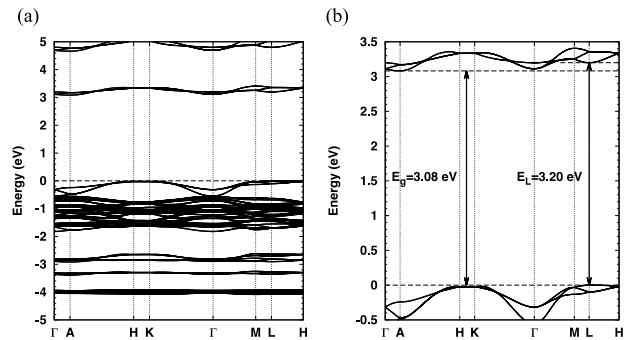

Figure 3 (a) Calculated band structure of $(MA)_2KBiCl_6$ (b) enlarged view of the band structure near the band gap. The following high symmetry points in the first Brillouin zone were used: Γ (0,0,0), A (0,0,0.5), H (-0.333,0.667,0.5), K (-0.333,0.667,0), M (0,0.5,0) and L (0,0.5,0.5). The band edges are found at L in the valence band and A in the conduction band.

Single crystal stiffness constants ($C_{ij}$) were calculated from the stress-strain relationship by applying 2 types of strain: $e_1$ and $e_3+e_4$ to the cell (each strain with ±0.5% and ±1% deformation) and relaxing the internal degrees of freedom.[39] The tensorial analysis[40] of the $C_{ij}$'s reveals the whole anisotropy elasticity of $(MA)_2KBiCl_6$ (Table S6). In order to better understand the mechanical properties of $(MA)_2KBiCl_6$, a 3D directional Young's modulus and its contour plots were calculated by tensorial analysis of the $C_{ij}$'s and is shown in Figure S9. It is easy to see that the directions along Bi-Cl-K bonds have the largest value of the Young's modulus, which is very similar to $(MA)PbX_3$ (X = Cl, Br and I).[30] The calculations predict a Young's modulus of 13.07 GPa normal to the (001) facet, which is in reasonable agreement with the experimental nanoindentation results.

Resonant ultrasound spectroscopy (RUS) provides a highly sensitive method for detecting phase transitions through the influence of strain coupling effects.[41] The spectra reported here were obtained using electronics designed by Dr A.Migliori in Los Alamos, with a maximum applied voltage of 2 V, and a sample holder described by McKnight et al.[42] which was lowered into an Oxford Instruments Teslatron cryostat. Each spectrum contained 135,000 data points in the frequency range 0.01-1 MHz collected from an irregular shaped crystal (maximum dimensions ~1.5-2 mm and mass 0.0012 g) during a sequence of cooling and heating between 307K and 1.5K in 5K steps. The sample chamber had first been evacuated and then filled with a few mbars of helium as exchange gas. A 10 minute settle time before data collection was allowed at each set point to allow thermal equilibration.

Figure 4 shows segments of the spectra stacked in proportion to the temperature at which they were collected. Four features are immediately apparent (i) there is a break in trend and peak widths at ~260 K; (ii) there is then a clear hysteresis between cooling and heating through the interval ~125-260K; (iii) the trend with reducing temperature for most resonances is of elastic softening below ~170 K, and (iv) individual resonances weaken or disappear in the vicinity of 50 K but reappear at the lowest temperatures. Selected peaks were fitted with an asymmetric Lorentzian function to follow their frequency, $f$, and width at half maximum height, $\Delta f$. Values of the elastic constants, which determine each resonant mode scale with $f^2$, and the inverse mechanical quality factor, given by $Q^{-1} = \Delta f/f$, are a measure of acoustic attenuation. Data from peaks with resonance frequencies ~ 400 and ~460 kHz at room temperature are shown in Figure S10. Stepwise softening immediately below ~265 K, accompanied by a peak in $Q^{-1}$, is similar to that seen at the octahedral tilting transition in $EuTiO_3$,[43] except that there is a hysteresis in the position of the anomaly between heating (267 K) and cooling (260 K). In comparison with $EuTiO_3$, the magnitude of the softening is small but the overall pattern is characteristic of a first order phase transition with weak coupling of strain to the driving order parameter. The softening trend with cooling below ~150 K is indicative of another incipient instability, as seen also in $LaAlO_3$,[44] and the near disappearance of resonance peaks in the interval ~30-60 K is suggestive of some additional process with weak strain coupling.

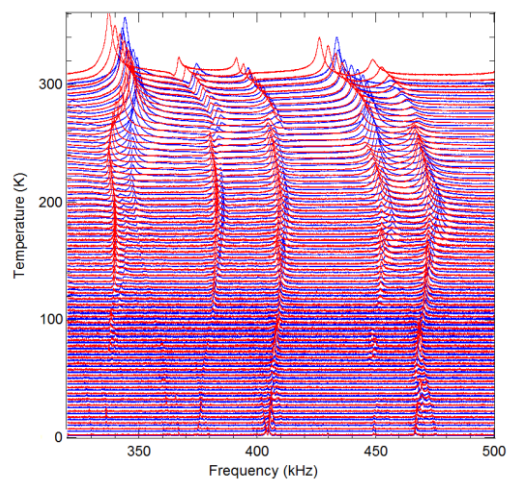

Figure 4. Segments of RUS spectra collected during cooling (blue) and heating (red). The y-axis is the amplitude in volts from the detecting transducer, but the spectra have been offset in proportion to the temperature at which they were collected and the axis labelled as temperature.

In conclusion, we have discovered a lead-free, hybrid double perovskite, $(MA)_2KBiCl_6$, which shows striking similarities to the widely-studied lead halide perovskites $(MA)PbX_3$. The system has a wide band gap of 3.04eV, which is similar to that of the compound $(MA)PbCl_3$ (~3.0eV).[45] Resonant ultrasound spectroscopy points to a series of phase transitions that may be associated with hydrogen bond ordering or octahedral tilting transitions, or both. Our findings also suggest that many other compositions should be explored with a view to making a lead-free, semiconducting analogue of $(MA)PbI_3$; these include $(MA)_2CuBiI_6$, $(MA)_2AgBiI_6$, and $(MA)_2TlBiI_6$. Further work along these lines is currently underway in our laboratory.

**Acknowledgements:** F. Wei is a holder of A*STAR international fellowship granted by Agency for Science, Technology and Research, Singapore. G. Kieslich and A.K. Cheetham thank the Ras al Kaihmah Center for Advanced Materials for financial support. S. Sun and Z. Deng would like to thank the Cambridge Overseas Trust and China Scholarship Council. The calculations were performed at the Cambridge HPCS and the UK National Supercomputing Service, ARCHER. Access to the latter was obtained via the UKCP consortium and funded by EPSRC under Grant No. EP/K014560/1. All necessary computational data is included in the manuscript or supplementary information. RUS facilities were established in Cambridge through grants to MAC from the Natural Environment Research Council (NE/B505738/1, NE/F017081/1) and the Engineering and Physical Sciences Research Council (EP/I036079/1).

Supplementary information:

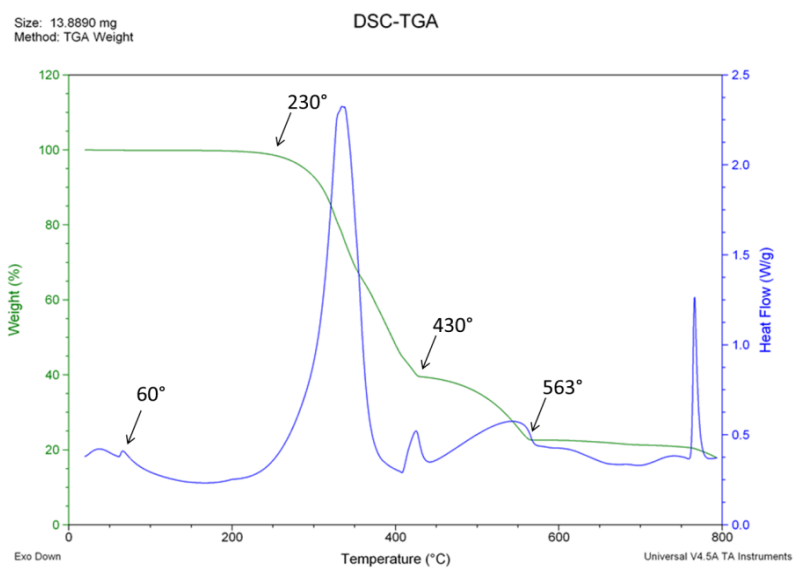

Figure S1. TGA and DSC curve for the (MA)$_2$KBiCl$_6$.

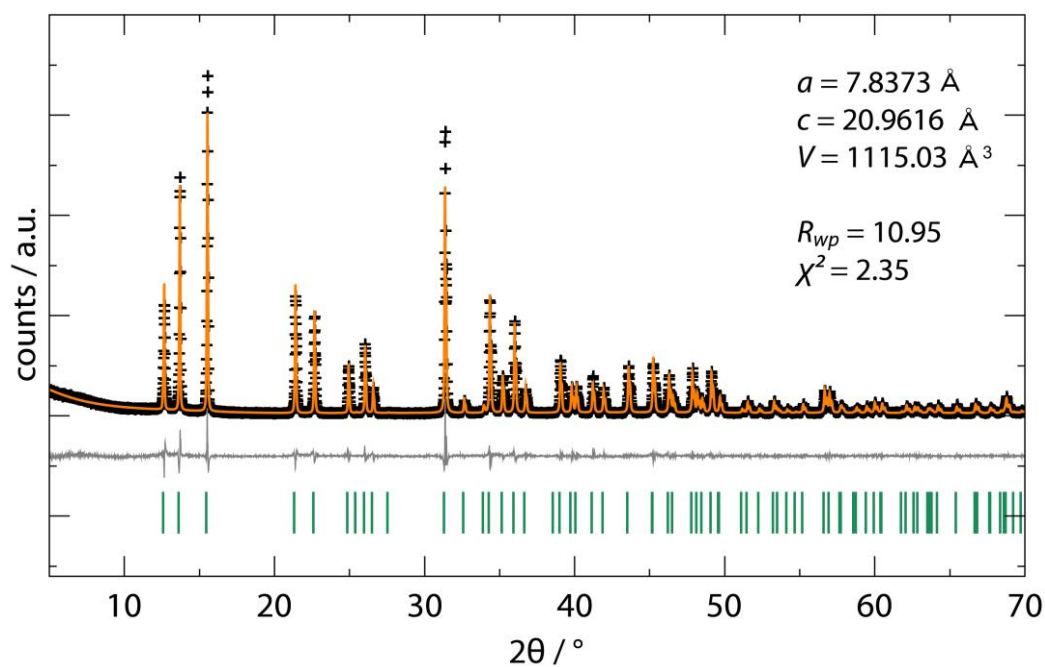

Figure S2. Powder X-ray diffraction pattern for MA$_2$KBiCl$_6$ and Pawley fit. The lattice parameter (inset) obtained from the Pawley fit agree well with lattice parameter from Single crystal X-ray diffraction. Blue - experimental, red – calculated, grey – differences between experimental and calculated curves. The tics were intensities are expected are shown in green.

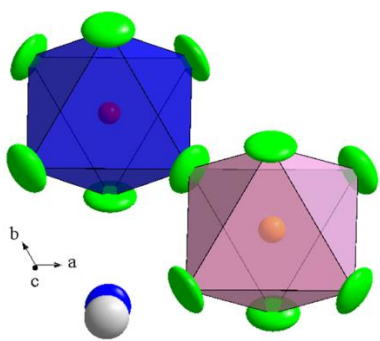

Figure S3. Ellipsoids for the Cl, C and N, viewed slightly away from *c* axis, drawn in 50% probability.

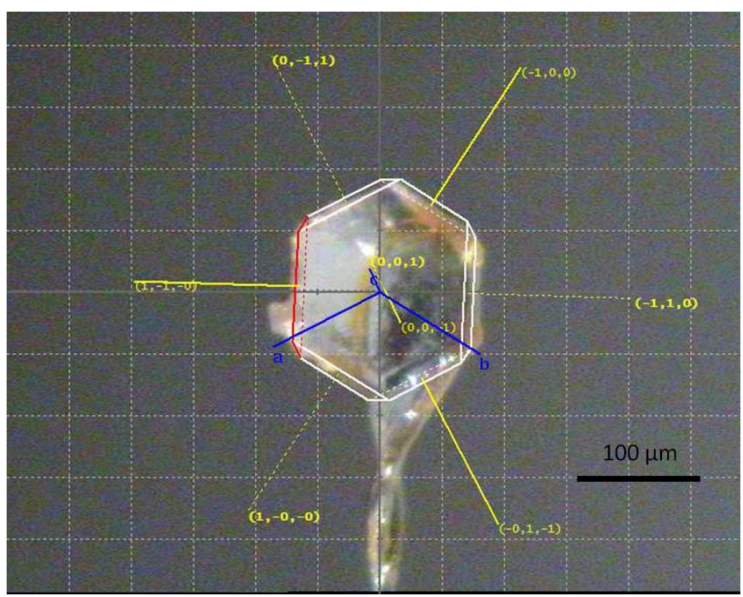

Figure S4. Photograph of (MA)$_2$KBiCl$_6$ with face indexing using single crystal X-ray diffraction.

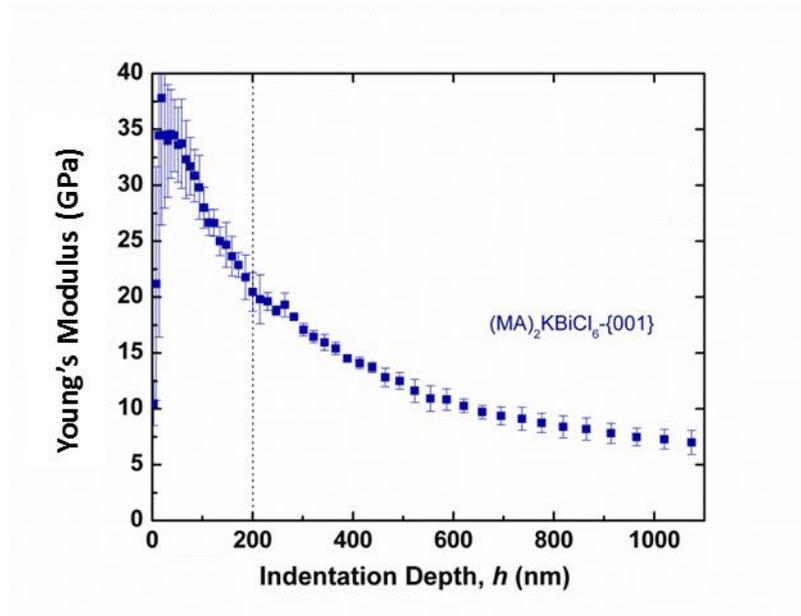

Figure S5. Young's modulus against indentation depth.

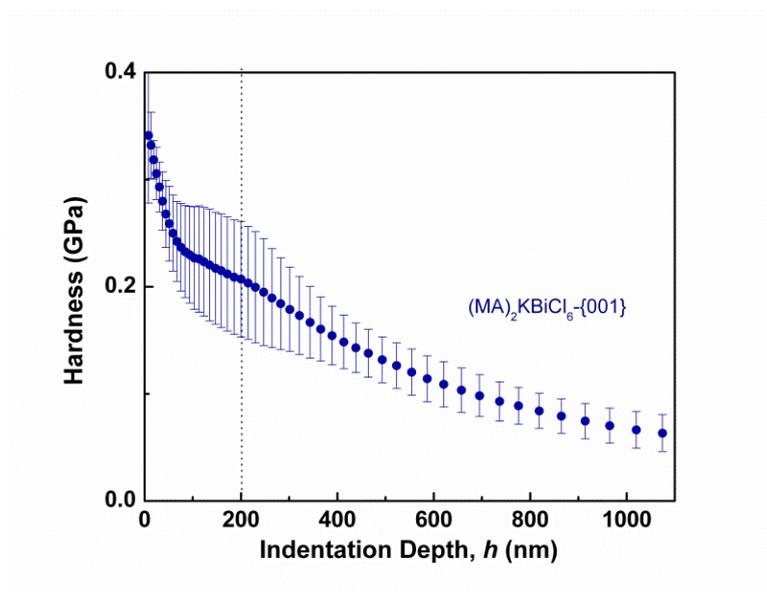

Figure S6. Hardness against indentation depth.

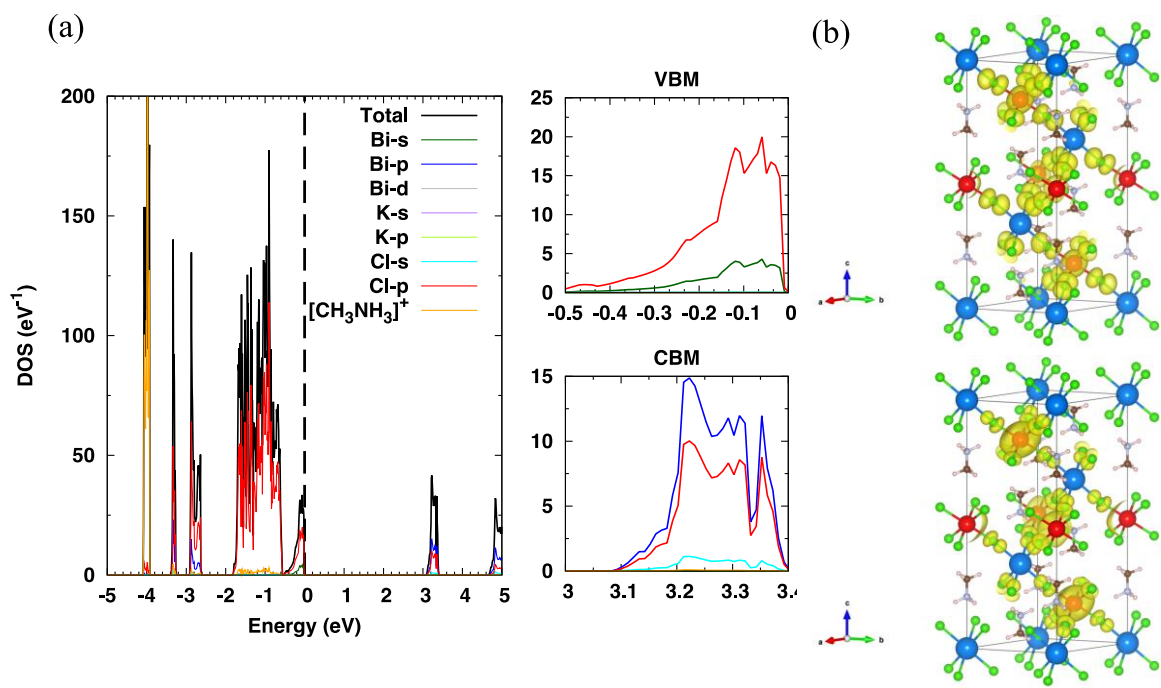

Figure S7 (a) Calculated total and partial density of states of (MA)$_2$KBiCl$_6$ (b) calculated band decomposed partial charge density (PCD) at the VBM (top) and the CBM (bottom).. The PCD is visualized using the VESTA program[1] and the electron isosurface level is set at 0.001 eV/Å$^3$. The density clouds are colored yellow and the atoms are: Bi-red, K-blue, Cl-green, C-brown, N-light blue and H-white.

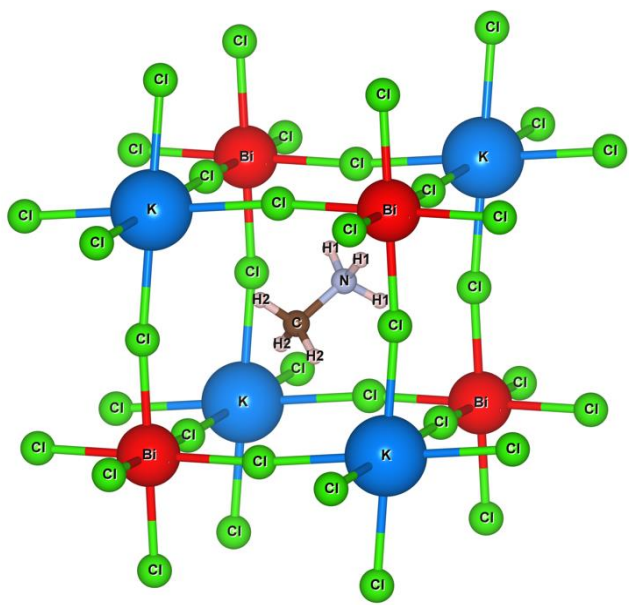

Figure S8. Atomic labels for (MA)$_2$KBiCl$_6$.

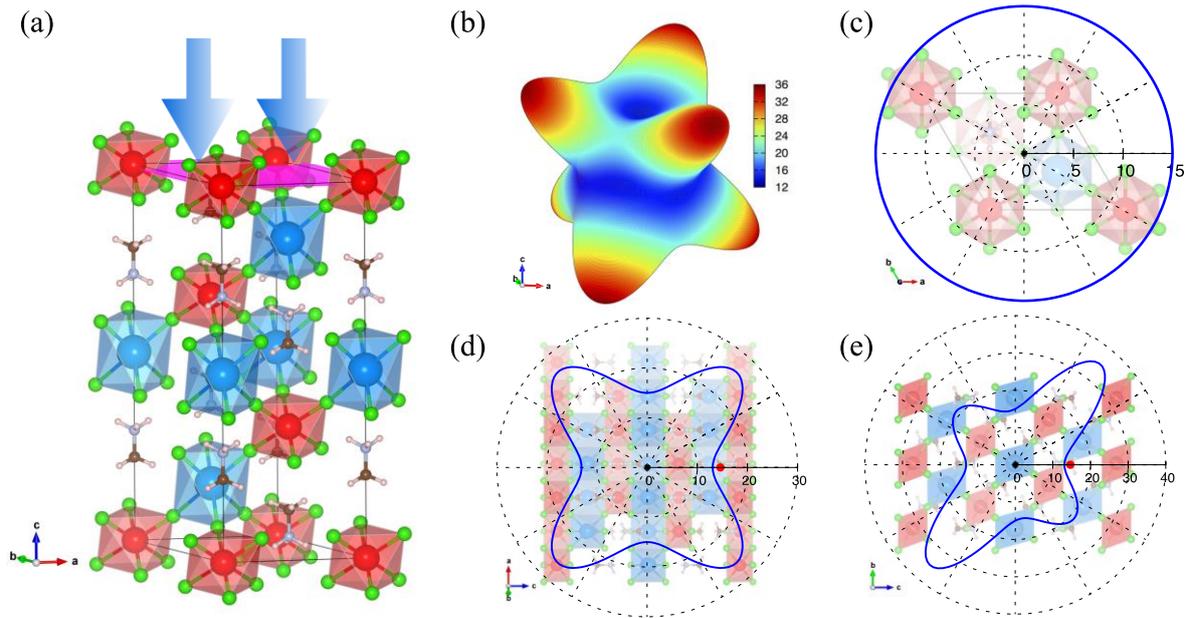

Figure S9 (a) Nanoindentation plane (purple) and direction (blue arrow) (b) Calculated 3D directional Young's modulus of $(MA)_2KBiCl_6$ and its contour plot on (c) (001) plane (d) (010) plane and (e) the plane perpendicular to [100]. Nano-indentation Young's modulus is shown as red dot in (d) and (e) along c-axis compared with DFT results (blue curve). Units shown are in GPa.

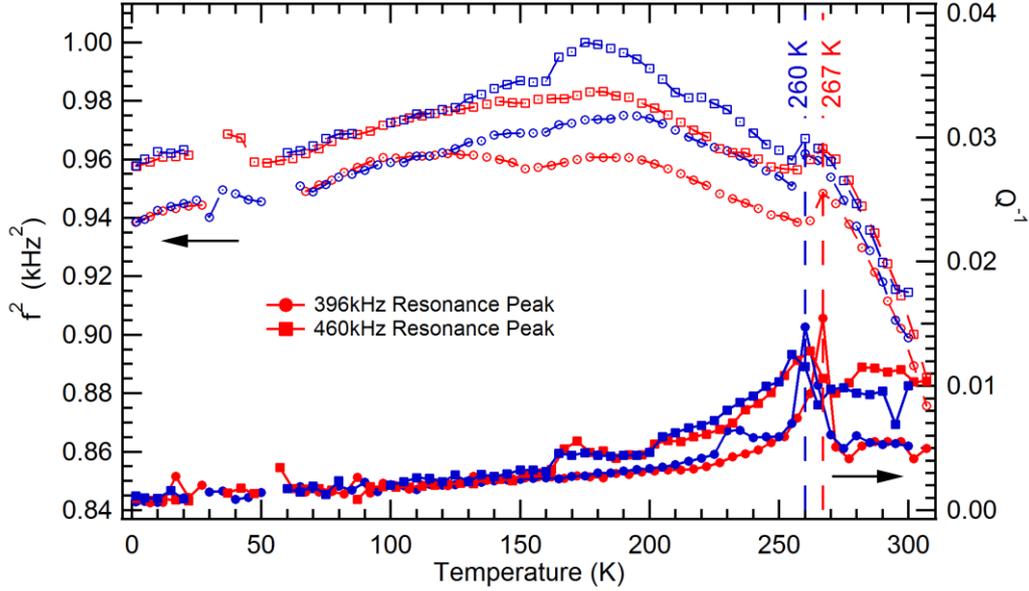

Figure S10. Variation of $f^2$ (proportional to elastic constants) and $Q^{-1}$ (indicative of acoustic attenuation) from selected resonance peaks in RUS spectra collected during cooling (blue) and heating (red). There is clear evidence of a phase transition at ~260 K, with a step in $f^2$ and peak in $Q^{-1}$ occurring at 260 K during cooling and 267 K during heating.

Table S1. Atomic coordinates for $(MA)_2KBiCl_6$ obtained from single crystal diffraction compared with values from DFT geometry optimization. Atomic labels are shown in Figure S8.

| Atom | Experiments | | | | DFT | | |
|---|---|---|---|---|---|---|---|
| | x | y | z | Biso | x | y | z |
| Bi | 0.6667 | 0.3333 | 0.3333 | 0.0294(2) | 0.66667 | 0.33333 | 0.33333 |
| K | 0.3333 | 0.6667 | 0.1667 | 0.0364(7) | 0.33333 | 0.66667 | 0.16667 |
| Cl | 0.5029(2) | 0.4971(2) | 0.2619(1) | 0.105(1) | 0.50064 | 0.49936 | 0.26155 |
| C | 0.3333 | 0.6667 | 0.4456(9) | 0.121(9) | 0.33333 | 0.66667 | 0.44560 |
| N | 0.3333 | 0.6667 | 0.3815(8) | 0.133(8) | 0.33333 | 0.66667 | 0.37450 |
| H1 | - | - | - | - | 0.25664 | 0.74336 | 0.46244 |
| H2 | - | - | - | - | 0.40409 | 0.59591 | 0.35508 |

Table S2. Anisotropic displacement parameters (Å$^2$) for (MA)$_2$KBiCl$_6$.

| Atom | U11 | U22 | U33 | U23 | U13 | U12 |
|------|-----|-----|-----|-----|-----|-----|
| Bi | 0.0269(2) | 0.0269(2) | 0.0345(3) | 0 | 0 | 0.01345(11) |
| K | 0.0390(11) | 0.0390(11) | 0.0312(15) | 0 | 0 | 0.0195(5) |
| Cl | 0.145(2) | 0.145(2) | 0.0849(16) | 0.0203(7) | -0.0203(7) | 0.118(3) |
| C | 0.153(15) | 0.153(15) | 0.057(10) | 0 | 0 | 0.077(7) |
| N | 0.165(13) | 0.165(13) | 0.070(10) | 0 | 0 | 0.082(6) |

Table S3. Interatomic distances for (MA)$_2$KBiCl$_6$ from experiments compared with DFT.

| Atom 1 | Atom 2 | D (Å) | D$_{DFT}$ (Å) |
|--------|--------|-------|---------------|
| Bi | Cl | 2.681(2) | 2.706 |
| K | Cl | 3.049(2) | 3.016 |
| N | C | 1.35(3) | 1.492 |
| C | H1 | - | 1.097 |
| N | H2 | - | 1.041 |
| H1 | Cl | - | 3.032 |
| H2 | Cl | - | 2.359 |
| N | Cl | 3.406(1) | 3.279 |
| C | Cl | 3.848(2) | 3.856 |

Table S4. Bond angles for (MA)$_2$KBiCl$_6$ from experiments compared with DFT.

| Bond Angle (°) | Exp | DFT |
|----------------|-----|-----|
| K-Cl-Bi | 173.04(12) | 172.51 |
| Cl$^1$-Bi-Cl$^2$ | 91.76(9) | 92.00 |
| Cl$^1$-Bi-Cl$^3$ | 88.24(9) | 88.00 |
| Cl$^5$-K1-Cl$^4$ | 98.33(9) | 98.86 |
| Cl$^4$-K1-Cl$^6$ | 81.67(9) | 81.14 |
| C-H1··Cl | - | 132.32 |
| N-H2··Cl | - | 146.70 |

Symmetry code: $^1$1-y, x-y, z; $^2$1+y, -x, z; $^3$ 4/3-x, 2/3-y, 2/3-z; $^4$ 2/3-x, 4/3-y, 1/3-z; $^5$ 1-y, 1+x-y, z; $^6$2/3-y+x, 1/3+x, 1/3-z.

Table S5. Experimental details.

| | |
|---|---|
| **Crystal data** | |
| Chemical formula | $C_2N_2H_{12}KBiCl_6$ |
| Chemical formula weight | 524.9 |
| Temperature (K) | 300 |
| Cell Setting | R |
| Superspace group | $R\bar{3}m$ |
| $a$ (Å) | 7.8372(2) |
| $c$ (Å) | 20.9938 (2) |
| Volume (Å$^3$) | 1116.72(7) |
| Formula units (Z) | 3 |
| $D_x$ (Mg m$^{-3}$) | 2.288 |
| Crystal size (mm) | 0.0809×0.1048 ×0.1384 |
| Crystal colour | Colorless |
| | |
| **Data collection** | |
| Diffractometer | |
| Radiation type | Mo $K\alpha$ |
| Wavelength (Å) | 0.71073 |
| Absorption correction type | Analytical |
| Absorption coefficient μ (mm$^{-1}$) | 13.158 |
| Range of h, k, l | -9≤h≤7, -7≤k≤9, -26≤l≤26 |
| No. of measured reflections | 1487 |
| No. of unique reflections | 316 |
| Criterion for observed reflections | $I > 2\sigma (I)$ |
| | |
| **Refinement** | |
| Refinement on | $F^2$ |
| $R, wR$ (all reflections) | 0.0226, 0.0540 |
| S | 1.144 |
| No. of parameters | 17 |
| Weighting scheme | $w = [\sigma^2(F) + (0.01F)^2]^{-1}$ |
| $(\Delta/s.u.)_{max}$ | 0.002 |
| $\Delta\rho_{max}$ (e Å$^{-3}$) | 0.65 |
| $\Delta\rho_{min}$ (e Å$^{-3}$) | -0.69 |
| Extinction correction | None |
| Source of atomic scattering factors | International Tables for Crystallography (1992, Vol.C)[2] |

Table S6. DFT calculated single crystal elastic stiffness constants ($C_{ij}$) and elastic properties of bulk modulus (B), shear modulus (G), Young's modulus (E) and Poisson's ratio (ν). Except ν, all the units are in GPa.

| Elastic Constants | $C_{11}$ | $C_{12}$ | $C_{13}$ | $C_{14}$ | $C_{33}$ | $C_{44}$ |
|---|---|---|---|---|---|---|
|  | 26.26 | 9.01 | 18.10 | -3.40 | 31.64 | 12.59 |
| Elastic Properties | B | G | E | ν |  |  |
|  | 18.05 | 4.46~14.54 | 13.07~20.22 | 0.12~0.69 |  |  |